\documentclass[runningheads,11pt]{llncs}

\usepackage{geometry}
\usepackage{amsfonts}
\usepackage[utf8]{inputenc}
\usepackage[T1]{fontenc}
\usepackage{lmodern}
\usepackage{cite}
\usepackage{spverbatim}
\usepackage{breakcites}
\usepackage{amssymb}
\usepackage{amsmath}
\usepackage[pdftex]{graphicx}
\usepackage{float}
\usepackage{hyperref}
\hypersetup{colorlinks,allcolors=black}
\usepackage{multicol,xparse}
\usepackage[american]{babel}
\usepackage[table]{xcolor}
\usepackage{hhline}
\usepackage{pifont}
\usepackage{authblk}

\usepackage{enumitem}

\makeatletter
\def\@fnsymbol#1{\ensuremath{\ifcase#1\or \dagger\or \ddagger\or
		\mathsection\or \mathparagraph\or \|\or **\or \dagger\dagger
		\or \ddagger\ddagger \else\@ctrerr\fi}}
\makeatother

\title{Certificate and Signature Free Anonymity for V2V Communications\thanks{This is the full version of the paper that appeared in 2017 IEEE Vehicular Networking Conference (VNC), pp. 139-146. DOI: {10.1109/VNC.2017.8275624}~\cite{Vipin[17]}}}
\author{Vipin Singh Sehrawat \inst{1} \and Yogendra Shah \inst{2} \and Vinod Kumar Choyi \inst{2} \and Alec Brusilovsky \inst{2} \and Samir Ferdi \inst{3}}

\institute{Department of Computer Science,\\
The University of Texas at Dallas, Richardson, USA\\
\email{\small \tt vipinsingh.sehrawat@utdallas.edu}
\and 
InterDigital Communications Inc., Conshohocken, USA\\
\email{\{\small \tt yogendra.shah, vinod.choyi, alec.brusilovsky\} @interdigital.com}
\and
InterDigital Canada Ltee, Montreal, Canada\\
\email{samir.ferdi@interdigital.com}}

\titlerunning{Certificate and Signature Free Anonymity for V2V Communications}
\authorrunning{V. S. Sehrawat el al.}

\begin{document}
\maketitle

\begin{abstract}
	\normalsize
Anonymity is a desirable feature for vehicle-to-vehicle (V2V) communications, but it conflicts with other requirements such as non-repudiation and revocation. Existing, pseudonym-based V2V communications schemes rely on certificate generation and signature verification. These schemes require cumbersome key management, frequent updating of certificate chains and other costly procedures such as cryptographic pairings. In this paper, we present novel V2V communications schemes, that provide authentication, authorization, anonymity, non-repudiation, replay protection, pseudonym revocation, and forward secrecy without relying on traditional certificate generation and signature verification. Security and privacy of our schemes rely on hard problems in number theory. Furthermore, our schemes guarantee security and privacy in the presence of subsets of colluding malicious parties, provided that the cardinality of such sets is below a fixed threshold.
\end{abstract}

\section{Introduction}
\label{sec:introd}
V2V communications is defined as the wireless transmission of data between motor vehicles. The National Highway Traffic Safety Administration (NHTSA) has made ensuring data security and privacy a high-priority in the proposed safety requirements for V2V communications~\cite{NHTSA[17]}. The requirements draw out the need to make it infeasible for a polynomial adversary to identify and track vehicles. A trusted third party (e.g. the service provider) should be able to identify and \textit{blacklist} malicious vehicles, but without compromising the vehicles' identity. Conflicting requirements such as anonymity and non-repudiation makes designing efficient V2V communications schemes a challenge. In this paper, we present provably secure, efficient V2V communications schemes, which do not rely on the traditional certificate generation and signature verification, and provide anonymity, non-repudiation, replay protection and pseudonym revocation. 

Numerous pseudonym-based V2V communications schemes have been proposed but these rely on certificate generation and signature verification. Petit et al.~\cite{Petit[15]} provide a comprehensive survey of the existing pseudonym-based V2V schemes. Such schemes carry the inherent drawbacks of certificate based pseudonym generation, such as key management, cryptographic pairings, frequent updates of the certificate chains, cumbersome revocation, etc. All existing schemes require a \textit{cluster head} to form and maintain multicast groups/clusters of vehicles. Thus, in order to guarantee security and privacy, the cluster head is assumed to be honest. There exist cryptographic solutions, that do not rely on certificates and achieve anonymity in more constrained models, but no such solution exists for V2V communications. 

Damgard~\cite{Dam[95]} employed general complexity theoretic primitives (one-way functions and zero-knowledge proofs) to solve the problem of allowing multiple users to anonymously transfer credentials from one organization to another. But that scheme does not protect against actively dishonest users and is impractical due to the costly underlying primitives. Lysyanskaya et al.~\cite{Anna[99]} developed a general credential system, which dealt with the case of users acting as active adversaries, but that scheme too uses one-way functions and zero knowledge proofs, making it impractical. If we expand the problem space to include revocation and multiple-use credentials along with (pseudo)anonymity, then the scheme by Camenisch et al.\cite{Jan[01]}, which is based on strong RSA assumption and the decisional Diffie-Hellman assumption, does satisfy the requirements. But in order to meet the security and privacy requirements, the freshness of the \textit{one-show credential} must be verified by the credential issuing organization. Moreover, that scheme is defined for an entirely different model to V2V communications. The model of that scheme defines mutliple users, that are issued certain credentials by different organizations, allowing the users to authenticate themselves anonymously to the organizations. To the best of our knowledge, the schemes presented in this paper are the first scalable non-certificate based solutions that provide authentication, authorization, anonymity, non-repudiation, replay protection and revocation in a V2V communications setting.\\[2mm] 
\textbf{Organization.}
The organization of the paper is as follows. Section 2 gives a brief background, describing the concepts and mechanisms used in our schemes. Section 3 discusses the threat model and Section 4 outlines the broad phases of our V2V communications schemes. Section 5 presents our first scheme, \textit{V2VAuth}, which supports broadcast. In Section 6, we present a multicast variant of V2VAuth, named \textit{V2VMulticast}. Unlike the existing schemes, our multicast scheme does not require a cluster head. It employs V2VAuth for the communications but isolates the cluster and protects the intra-cluster communications by using a temporal and random cluster membership token, which is updated after a fixed time interval. Section 7 gives the security proofs and compares our schemes' privacy with that of the existing solutions. In Section 8, we analyze the time complexities of our schemes and compare them with the other well known V2V communications schemes, in particular the SCMS~\cite{Whyte[13]} scheme, which is currently being adopted for the DSRC (Dedicated Short Range Communications) standard. Section 9 discusses some possible directions for future work and Section 10 gives the conclusion. 

\section{Background}
\label{sec:back}
In this section, we recall the cryptographic primitives and results that are required for the rest of the paper. 

\subsection{Polynomial Interpolation}
Polynomial interpolation is the process of approximating a function $f$ by a polynomial $P$ via forcing it to have the same values as $f$ at a number of points. If we have $n+1$ distinct points, $\{x_i\}^{n}_{i=0}$, scattered throughout an interval [a,b], over which the function $f$ is defined, then we can find a polynomial of degree $n$ with the same values as $f$ at the inputs $\{x_i\}^{n}_{i=0}$. To construct a polynomial of degree $n$ passing through $n+1$ data points $(x_0 , y_0), (x_1 , y_1), ... , (x_n, y_n)$, we begin by constructing a set of basis polynomials, $L_{n,k}(x_j)$ $(0 \leq j \leq n)$, such that:
\begin{equation}
    L_{n,k} = 
    \begin{cases}
      1, & \text{if}\ j = k\\
      0, & \text{otherwise}
    \end{cases}
    \notag
  \end{equation}
Once we have constructed the basis polynomials, we can form the $n^{th}$ degree Lagrange interpolating polynomial as: $L(x) = \Sigma_{k=0}^{n} L_{n,k}(x)$. When $x = x_j$, every basis function vanishes, except for $L_{n,j}(x)$, which has value $1$. Thus, $L(x_j) = y_j$. 
    
\subsection{Secret Sharing}
A secret sharing scheme~\cite{Shamir[79],Blakley[79],Ito[87]} is a method by which a dealer, holding a secret string, distributes strings, called shares, to parties such that authorized subsets of parties, specified by a public access structure, can reconstruct the secret. Secret sharing is the foundation of multiple cryptographic tools (in addition to its obvious use in secure storage), such as, threhsold cryptography~\cite{Yvo[89]}, secure multiparty computation~\cite{Micali[91]}, attribute-based encryption~\cite{Goyal[06]}, generalized oblivious transfer~\cite{Tassa[11]}, perfectly secure message transmission~\cite{Danny[93]}, e-voting~\cite{Berry[99],Yung[04]} and e-auctions~\cite{Mic[98],Peter[09]}. The extensive survey by Beimel~\cite{Beimel[11]} gives a review of the notable results in the area. 

\begin{definition}[Secret Sharing]\label{def.1}
	\emph{A secret sharing scheme with respect to an access structure $\Gamma$ a set of $\ell$ polynomial-time parties $\mathcal{P} = \{P_1, \dots, P_\ell \}$, and a set of secrets $\mathcal{K}$, consists of a pair of polynomial-time algorithms, {\fontfamily{cmtt}\selectfont (Share,Recon)}, where: 
		\begin{itemize}
			\item {\fontfamily{cmtt}\selectfont Share} is a randomized algorithm that gets a secret $k \in \mathcal{K}$ and access structure $\Gamma$ as inputs, and outputs $\ell$ shares, $\{\mathrm{\Pi}^{(k)}_1, \dots, \mathrm{\Pi}^{(k)}_\ell\},$ of $k$,
			\item {\fontfamily{cmtt}\selectfont Recon} is a deterministic algorithm that gets as input the shares of a subset $\mathcal{A} \subseteq \mathcal{P}$, denoted by $\{\mathrm{\Pi}^{(k)}_i\}_{i \in \mathcal{A}}$, and outputs a string in $\mathcal{K}$,
		\end{itemize}
		such that, the following two requirements are satisfied:
		\begin{enumerate}
			\item \textit{Perfect Correctness:} for all secrets $k \in \mathcal{K}$ and every authorized subset $\mathcal{A} \in \Gamma$, it holds that: \\ Pr[{\fontfamily{cmtt}\selectfont Recon}$(\{\mathrm{\Pi}^{(k)}_i\}_{i \in \mathcal{A}}, \mathcal{A}) = k] = 1,$ 
			\item \textit{Perfect Secrecy:} for every unauthorized subset $\mathcal{B} \notin \Gamma$ and all different secrets $k_1, k_2 \in \mathcal{K}$, it holds that the distributions $\{\mathrm{\Pi}_i^{(k_1)}\}_{i \in \mathcal{B}}$ and $\{\mathrm{\Pi}_i^{(k_2)}\}_{i \in \mathcal{B}}$ are indistinguishable.
	\end{enumerate}}
\end{definition}

\subsection{l-wise Independence via Bivariate Polynomials}
Let $\mathcal{F}$ be a function family over a finite field $\mathcal{H}$. The value of $f(h)$, is defined as $f(h) = P(0, h)$, where $P(x, y)$ is a bivariate polynomial. A provably secure variant of Shamir secret sharing allows the parties to evaluate the function $f$ by performing polynomial interpolation over their shares of the bivariate polynomial, $P(x,y)$\cite{Naor[99]}. If the degree of $x$ and $y$ in $P(x,y)$ is $d-1$ and $l-1$, respectively, then the scheme requires at least $d$ shares, $\{ \langle i_j, P(i_j,h) \rangle \}_{j=1}^d$, in order to perform polynomial interpolation and compute the free coefficient of the polynomial $P(\cdot, h)$, namely the value $f(h) = P(0,h)$. The scheme allows $l$ secure function evaluations.

\subsection{Bivariate Polynomial Modulo RSA Composite}
Boneh et al.\cite{Boneh[14]} proved that low degree bivariate polynomials can define collision and second preimage resistant functions of the form, $f : \mathbb{Z}_N \times \mathbb{Z}_N \rightarrow \mathbb{Z}_N$, where $N$ is a random RSA modulus of secret factorization. In our schemes, such polynomials play a central role in pseudonym generation and authentication. 
    
\subsection{Quadratic Residue}
An integer \textit{x} is said to be a Quadratic Residue (QR) modulo \textit{n} if there exists an integer \textit{y} such that: $x = y^{2} \bmod ~n$.
If $n$ is a prime then the quadratic residuosity of $x$ is computed using the \textit{Legendre symbol}, which is defined as:
\begin{equation}
\left( \dfrac{x}{n} \right) =
\begin{cases}
      1, & \text{if $x$ is a QR modulo $n$}\\
      -1, & \text{if $x$ is not a QR modulo $n$}\\
      0, & \text{if} \: x = 0 \bmod n\\
    \end{cases}
    \notag
    \end{equation}
Legendre symbol is computed using Fermat's little theorem as:  

\begin{center} $\left( \dfrac{x}{n} \right) = x^{\frac{n-1}{2}} \bmod n$\end{center}   

If $n = \Pi^{t}_{i=1} p_i$, where $p_i \: (1 \leq i \leq t)$ are primes, then it follows from Chinese Remainder Theorem that $x$ is a quadratic residue modulo $n$ if and only if it is a quadratic residue w.r.t. $p_i \: (1 \leq i \leq t)$. Hence, without knowing all primes factors of $n$, there is no way to definitively say whether a given integer is a QR modulo $n$ or not. Legendre symbol, when computed modulo a composite is termed Jacobi symbol, and is defined as the product of $x$'s Legendre symbol w.r.t. to each prime factor of $n$, i.e.,       
\begin{center}
$\left( \dfrac{x}{n} \right) = \left(\dfrac{x}{p_1}\right) \left(\dfrac{x}{p_2}\right) \dots \left(\dfrac{x}{p_t}\right)$
\end{center}
Therefore, quadratic residuosity problem reduces to prime factorization. The number of quadratic residues and quadratic non residues in $\mathbb{Z}_p^*$ ($p$ is an odd prime) is $\frac{(p+1)}{2}$ and $\frac{(p-1)}{2}$, respectively. Quadratic residues modulo $n$ form a subgroup of $\mathbb{Z}^*_n$. The probability of $x \in \mathbb{QR}_n$ does not change if $x$ is a prime or a coprime w.r.t. $n$. The only exception is when $x = 0 \bmod n$, as then it is naturally a quadratic residue, with the Legendre symbol being $0$.

\subsection{Group Hash Function}
A group hash function $H: \{0,1\}^{*} \rightarrow G$, consists of two polynomial time algorithms, $HGen(1^k)$, which outputs a key $hk$, and $HEval(hk, X)$ ($X \in \{0, 1\}^k$), that deterministically outputs an image $H(hk, X) \in G$. 

\section{Threat Model}
\label{sec:thrt}
For an anonmymity scheme, C. Diaz~\cite{Diaz[05]} classified the adversary based on its properties and abilities as:
\begin{itemize}
\item Passive vs Active: A passive attacker listens to the communication and/or reads internal information of entities participating in the protocols, passive attackers typically
perform traffic analysis of the communication. Active attackers can add, remove or modify messages and adapt internal information of participating entities.
\item Internal vs External: An internal attacker controls one or several entities that are part of the system (e.g., the attacker controls communication nodes). External attackers
only control communication links.
\item Partial vs Global: A global attacker has access to the entire communication system (e.g., all communication links), while a partial attacker (also called local attacker in the literature) only sees part of the resources (e.g., a limited number of peers in a peer-to-peer network).
\item Static vs Adaptive: Static attackers control a predefined set of resources and are unable to alter their behavior once a transaction is in progress. Adaptive attackers gain control on new resources or modify their behavior, depending on intermediate results of the attack.
\item Temporary vs Permanent: Permanent adversary have been observing the system since it started functioning and knows its whole history. Temporary attackers start observing
or attacking the system at time $t_0$, and they do not have information on events previous to $t_0$.
\end{itemize}
In this paper, we consider the most powerful type of adversary, i.e., an \textit{Active Internal/External Global Adaptive} adversary. The goal of the adversary is to break the privacy of the scheme by either linking the pseudonyms belonging to the same vehicle or inverting the pseudonyms to retrieve vehicles' real identities.  

\begin{figure}[t!]
    \centering
    \includegraphics[scale=.5]{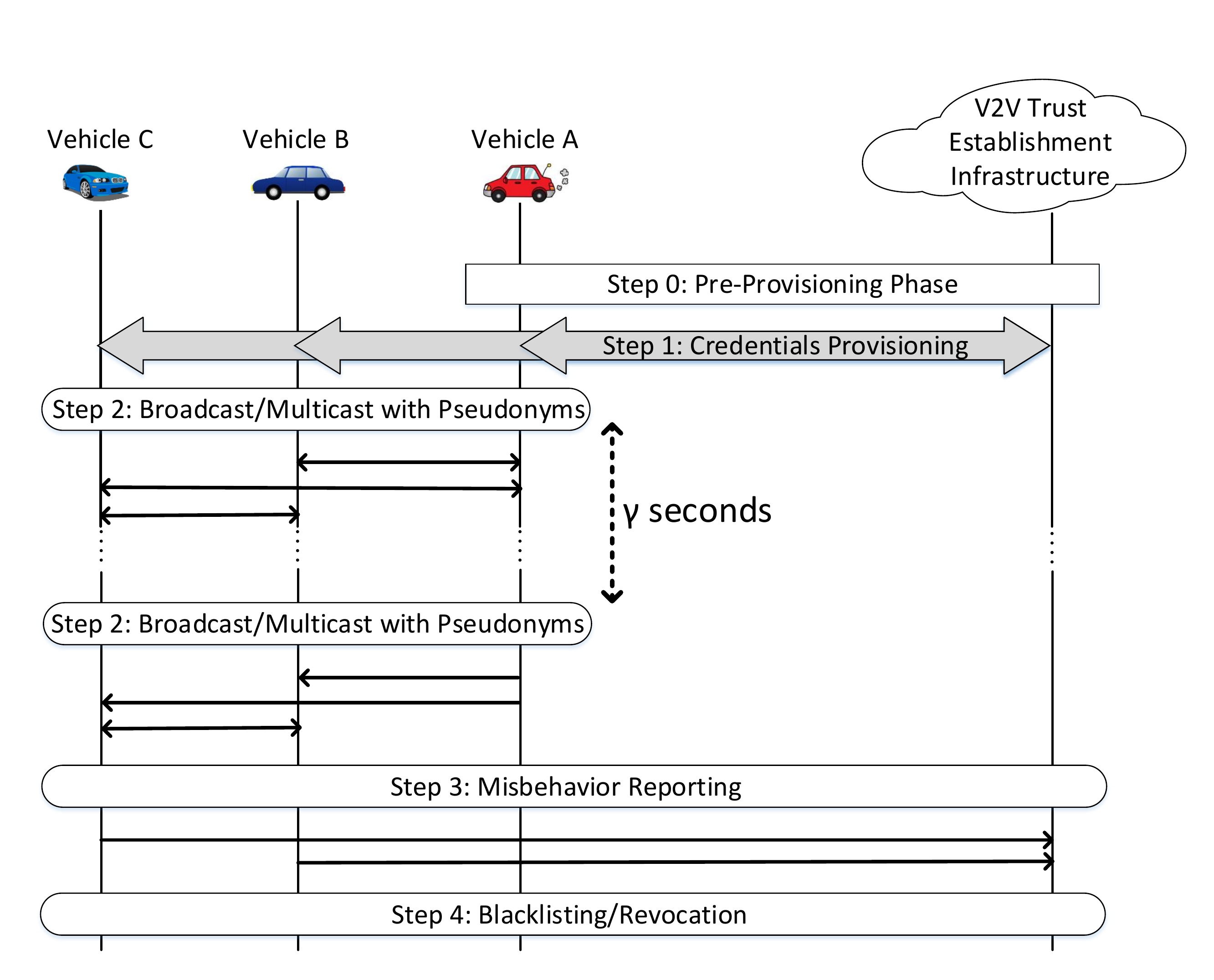}
    \caption{High level description of V2V Communications}
\end{figure}

\section{Phases of V2V Communications Scheme}
\label{sec:reqr}
In this section, we outline the broad phases of our V2V communications schemes. A V2V communications service provider either implements a V2V Trust Establishment Infrastructure (VTEI), that includes an Authentication Authority (AA) or uses the services of an AA from a trusted third party. The VTEI is responsible for provisioning appropriate policies associated with credential management, secure communications, misbehavior reporting and revocation. The following text explains the V2V communications phases depicted in Fig. 1. \\[1.2mm]
\textit{Step 0 (Pre-provisioning):} During this phase, the vehicles are provisioned with the AA's credentials (e.g. a public key). This step may be performed by the vehicle manufacturer in coordination with the transportation authority. The trust anchor's credentials are pre-provisioned in the vehicle's On-Board Unit (OBU). It is assumed that the long-term credentials are stored in a secure manner.\\[0.8mm]
\textit{Step 1 (Credentials Provisioning):} During this phase, the AA authenticates the vehicles in the network, and provisions unique credentials and associated parameters to each vehicle.\\[0.8mm]
\textit{Step 2 (Broadcast/Multicast with Pseudonyms):} Provisioned with the secret credentials and parameters, the vehicles generate pseudonyms and communicate anonymously.\\[0.8mm]
\textit{Step 3 (Misbehavior Reporting):} Vehicles may report malicious behavior of other vehicles. The VTEI is responsible for identifying the misbehaving vehicles from their pseudonyms.\\[0.8mm]
\textit{Step 4 (Blacklisting/Revocation):} Based on the misbehavior reports, the VTEI may determine the actual identity of the malicious vehicles and blacklist them. For each malicious vehicle, the VTEI broadcasts a parameter, enabling all vehicles in the network to identify messages from the malicious party. Backward privacy of the malicious vehicles is preserved and their real identities are never made public. The VTEI may whitelist previously blacklisted vehicles and allow them to rejoin the network by issuing fresh parameters. 

\section{Broadcast Scheme}
\label{sec:V2V}
In this section, we present a novel V2V communications scheme, named \textit{V2VAuth}, which supports broadcast. To the best of our knowledge, V2VAuth is the first V2V/V2X communication scheme that provides anonymity, nonrepudiation, authentication and revocation, without requiring certificate generation and signature verification. Unlike the existing pseudonym based V2V communications schemes, that require the vehicles to have the same pseudonym for some fixed time period, V2VAuth generates different, pseudorandom pseudonyms for each unique $<$vehicle, message, time frame$>$ triplet. Hence, V2VAuth provides better privacy than the existing solutions. V2VAuth is secure against both external and internal active adversary.\\[1.2mm]
\textbf{A note about notations:} for the sake of readability, we have taken an unusual approach towards notations in this paper. Due to the high number of variables used in our schemes, we have adopted an intuitive but unconventional naming methodology. The following notations are used throughout the rest of the paper.
\begin{itemize}
	\item $PRP:$ Pseudo random permutation (eg. AES), 
	\item $Pu_x/Pr_x:$ Public/Private key of party $x$, 
	\item $\mathbb{QR}_n:$ Subgroup of quadratic residues modulo $n$,
	\item $\mathbb{Z}_n:$ Group of integers modulo $n$, 
	\item $pid_{v_i}$ Permanent ID of vehicle $v_i$, 
	\item $tid_{v_i}$ Temporary ID of vehicle $v_i$. 
\end{itemize}
\textbf{Assumptions.} Our schemes operate under the following assumptions:
\begin{enumerate}
\item We assume that the size of a set of colluding dishonest vehicles, belonging to different vehicle classes, cannot exceed a threshold, $n'$.
\item The central authority cannot be impersonated and is reachable to all vehicles for registration, provisioning and malicious message reporting.  
\end{enumerate}

\subsection{Provisioning and Message Broadcasting}
The AA generates a secret key $k$ and decides the number of vehicle classes, $u$. AA generates $M = \Pi_{i=1}^j p_i$, where $p_1, p_2, \dots, p_j$, ($j \geq u$) are primes, such that $|M| \geq 2048$ bits. For each vehicle class $c_i$, a set, $\mathbb{P}_{c_i}$, of the prime factors of $M$, such that $\mathbb{P}_{c_i} \geq 1$. The size of each $\mathbb{P}_{c_i}$ set is the same and the no two sets overlap, i.e., $\forall i \neq j$, $\mathbb{P}_{c_i} \cap \mathbb{P}_{c_j} = \emptyset$. The size of these sets governs the success probability of the adversary. Thus, according to the security and privacy goals, and other factors such as traffic density, the AA decides the cardinality of these sets. W.l.o.g. let $|\mathbb{P}_{c}| = b$. A collision resistant bivariate polynomial, $P(x,y)$, defining function $f: \mathbb{QR}_M \times \mathbb{Z}_M \rightarrow \mathbb{QR}_M$, is generated, and a secure group hash function, $H: \{0, 1\}^* \longrightarrow \mathbb{Z}_M$ is fixed. Let the maximum degree of $x$ and $y$ in $P(x,y)$ be $d-1$ and $q-1$, respectively. Unless stated otherwise, all operations are performed modulo $M$.

\begin{figure}[t!]
\centering
  \includegraphics[scale=.7]{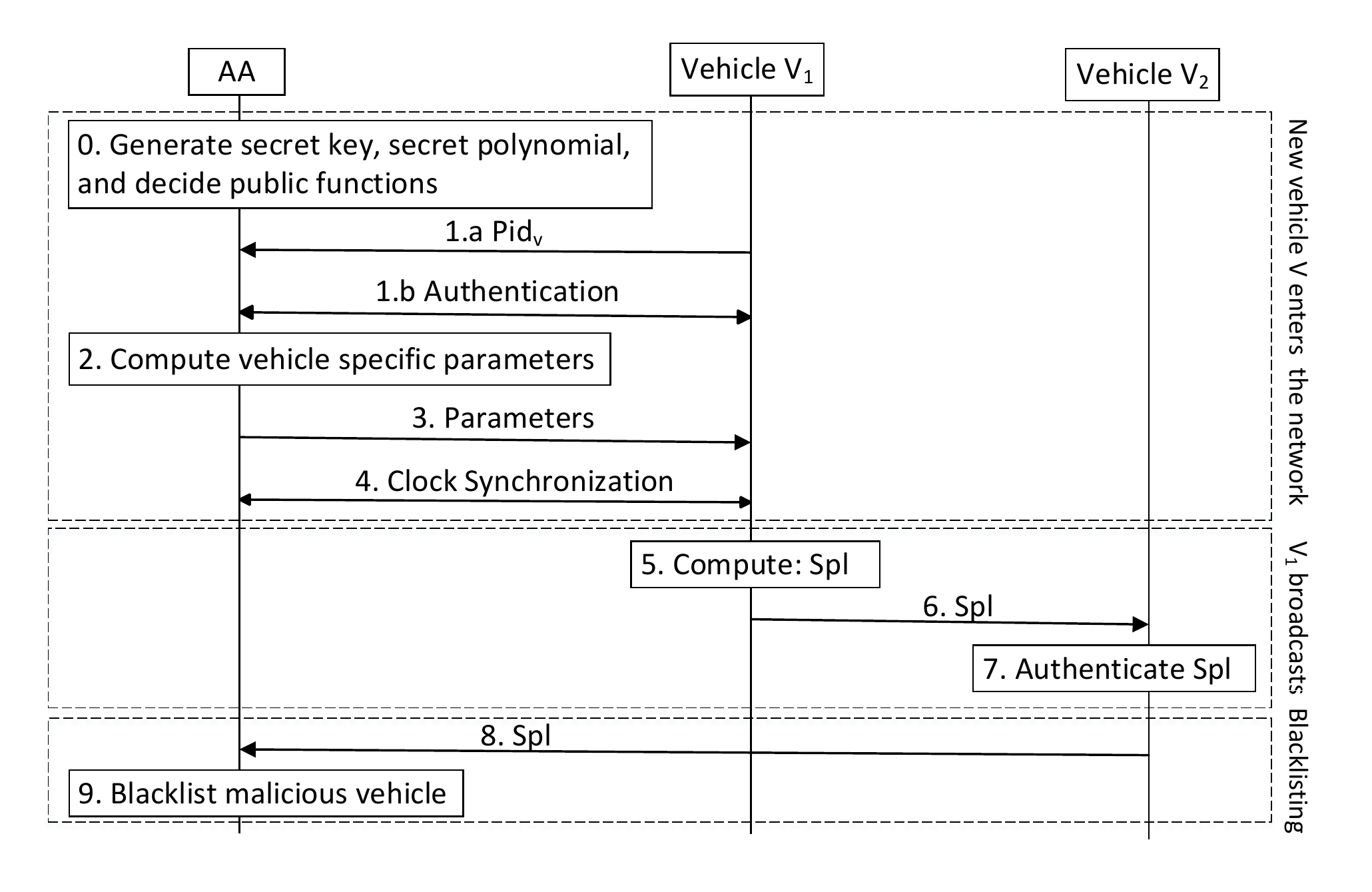}
	\caption{V2VAuth}
\end{figure}

AA maintains a public list, named $bad$, containing temporary ids of all blacklisted vehicles. A pseudonym generated by a blacklisted vehicle can be detected by using a public \textit{identification} function, $idnt(P_{(eno, v_i)}, tid_{v_j}, eno)$, which takes three inputs: a pseudonym, $P_{(eno, v_i)}$, temporary id ($tid_{v_j} \in bad$) of a blacklisted vehicle and a special hash, $eno$, of the message. Fig. 2 illustrates the high level call flow of \textit{V2VAuth} and the following text explains the scheme in detail.
\begin{enumerate}
\setcounter{enumi}{-1}
\item AA generates its secret parameters as:  
\begin{enumerate}[label=(\alph*)]
\item Generate three secret integers, $t$, $\alpha$ and $\gamma$. After every $\gamma$ seconds, $t$ gets incremented by $\alpha$. 
\item As AA knows the prime factorization of $M$, it generates the subgroup $\mathbb{QR}_M$. 
\end{enumerate}
\item Vehicle $v_1$ enters the network and requests authentication by sending its permanent id, $pid_{v_1}$, to the AA. It is assumed that a secure connection is established between $v_1$ and AA.  The AA authenticates $v_1$ by using some preselected authentication mechanism. 
\item Once $v_1$ is authenticated, the AA randomly decides a membership class, $c_1$, for $v_1$ and computes the vehicle specific parameters as:
\begin{enumerate}[label=(\alph*)]
\item Arrange the elements of $\mathbb{QR}_M$ in ascending order, let the ordered set be $\mathcal{QR}_M$. Compute $qr_{v_1} = PRP(k, pid_{v_1})$, and use it as an index in the set $\mathcal{QR}_M$ to select a quadratic residue as the temporary id. Hence, vehicle $v_1$'s temporary id is generated as: $tid_{v_1} = \mathcal{QR}_M[qr_{v_1}]$.  
\item Compute $P_{v_1} = P(tid_{v_1}, \cdot)$, which is a univariate polynomial of degree $q-1$ in $y$. 
\end{enumerate}
\item AA sends $(t, \alpha, \gamma, P_{v_1}, \mathbb{P}_{c_1})$ to $v_1$. 
\item Vehicle $v_1$ synchronizes its V2V clock (a clock used for V2V communications) with that of the AA.  
\item To broadcast a message $m$, $v_1$ generates a set $Spl$ as:
\begin{enumerate}[label=(\alph*)]
\item Use the group hash function $H$ to compute $eno = H(t \oplus m)$.
\item Evaluate $v_1$'s share of the secret polynomial at $eno$ to generate the pseudonym as: $P_{(eno, v_1)} = P_{v_1}(eno)$. 
\item Compute $vno = H(t \oplus m \oplus P_{(eno, v_1)})$.
\end{enumerate}
\item Vehicle $v_1$ broadcasts $Spl = \{P_{(eno, v_1)}, vno\}$ along with message $m$. 
\item On receiving the broadcast, vehicle $v_2$, belonging to class $c_2$, authenticates $Spl$ via the following procedure: 
\begin{enumerate}[label=(\alph*)]
\item First, verify that the message sender is not a blacklisted vehicle by testing the pseudonym $P_{(eno, v_1)}$ via the identification function, $idnt$. The procedure is explained in Section~\ref{Sec}.
\item\label{Step} Compute $eno = H(t \oplus m)$, $vno' = H(t \oplus m \oplus P_{(eno, v_1)})$, and verify that $vno = vno'$. This establishes the message and pseudonym integrity, and also verifies that $v_1$ holds the broadcast parameters, $(t$, $\alpha$, $\gamma)$. Thus, it serves a \textit{proof of possession} of the broadcast secrets.  
\item Next, authenticate the pseudonym by computing the Legendre symbols, $\left( \frac{P_{(eno, v_1)}}{p_{i}} \right)$, $\forall p_{i} \in \mathbb{P}_{c_2}$, and verifying that the pseudonym is a quadratic residue w.r.t. to each prime $p_i$. The probability of a fake pseudonym successfully passing the quadratic residue test of vehicle $v_2$ is $2^{-b}$, where $|\mathbb{P}_{c_2}| = b$. As the message is broadcast in the network, members of other classes also receive it. If members of $cl$ different classes receive the message, then the probability of a fake pseudonym getting through becomes $2^{-(cl+b)}$. AA can control this probability by modifying the size of the sets $\mathbb{P}_{c_i}$, changing the number of classes and setting a low upper bound on the maximum number of members of each class. \vspace{0.5mm}
\end{enumerate}
\textbf{Malicious Messages Handling and Blacklisting} 
\item To report a malicious message, $m$, the set $Spl$, accompanying $m$, is sent to the AA. 
\item The AA may identify and blacklist the sender as:
\begin{enumerate}[label=(\alph*)]
\item Compute $P' = P(\cdot, eno)$, which is a polynomial of degree $d-1$ in $x$. As AA knows the secret coefficients in polynomial $P(x, y)$ and the prime factorization of $M$, it solves $P' = P_{(eno, v_i)}$ for $x$, i.e. the $tid_{v_i}$ value of the malicious vehicle. We know that if the maximum $x$ degree in $P(x,y)$ is $d-1$, then the maximum number of real roots of $P' = P_{(eno, v_i)}$ is $d-1$. In order to identify the correct $tid_{v_i}$ value, AA performs the quadratic residuosity test modulo $M$ on each integer root of $P'$.\\ 
\underline{Claim.} We claim that one and only one root of $P' = P_{(eno, v_i)}$ can be a quadratic residue modulo $M$.\\
\underline{Proof.} We know that our bivariate polynomial, $P(x,y)$ is collision resistant i.e. no polynomial time algorithm can, with non-negligible advantage, generate $((\bar{x}, x) \in \mathbb{QR}_M)$ such that for $(y \in Z_M)$, $P(x,y) = P(\bar{x}, y)$. Therefore, out of the possible $d-1$ integer roots of $P' = P_{(eno, v_i)}$, only one belongs to the group $\mathbb{QR}_M$.   
\item Add $tid_{v_1}$ to the list of blacklisted vehicles. 
\end{enumerate}
\end{enumerate}
\subsection{Identifying Messages from Blacklisted Vehicles}\label{Sec}
Vehicles can identify the pseudonyms generated by the blacklisted vehicles via the identification function, $idnt$. Valid members of the broadcast network can evaluate $idnt$ via the following procedure:
\begin{enumerate}
\item First remove the `only y' terms (the terms without the variable $x$) from the pseudonym. This is achieved by evaluating the receiver's share of the secret polynomial, $P(x,y)$, at the hash ($eno$) of the received message. For example, let:\\ $P_{(eno, v_i)} = a \cdot tid_{v_i}^2 \cdot eno^8 + b \cdot tid_{v_i} \cdot eno^5 + c \cdot eno^3 + d \cdot tid_{v_i} \cdot eno^2 + e \cdot eno^4 + f \cdot eno$.\\ The receiver uses its share of the secret polynomial and the $eno$ value of the received message to compute:\\  $P_{v_i, eno} = P_{(eno, v_i)} - (c \cdot eno^3 + e \cdot eno^4 + f \cdot eno) = (a \cdot tid_{v_i}^2 \cdot eno^8 + b \cdot tid_{v_i} \cdot eno^5 + d \cdot tid_{v_i} \cdot eno^2)$.
\item Compute $\forall \, tid_{v_j} \in bad$, $Test_j = P_{v_i, y} \bmod tid_{v_j}$. If $Test_j = 0$, then the message was sent by the blacklisted vehicle with temporary id $tid_{v_j}$.  
\end{enumerate}

\subsection{Rejoining the Network}
The AA may allow a blacklisted vehicle, $v_a$, to rejoin the network. For example, a malware infected vehicle may be allowed to reenter after disinfection. In order to generate valid pseudonyms, $v_a$ must possess the current broadcast secrets, $t$, $\alpha$ and $\gamma$. In addition, $v_a$ requires a unique share of the secret polynomial, $P(x,y)$, as the pseudonyms that can be generated via its previous share, $P_{v_a}$, were blacklisted by the AA. In order to generate a fresh, unique share of $P(x,y)$, AA encrypts its secret key $k$ with itself to generate $k' = PRP(k, k)$. The new index, which decides the temporary id, $tid_{v_a}$, is generated as: $qr'_a = PRP(k', tid_{v_b})$. The rest of the V2VAuth procedure remains the same.   

\subsection{Replay Protection}
Each vehicle saves all $vno$ values received till the next update of the broadcast secret $t$. In the meantime, if a duplicate message with the same $vno$ value is received then it is rejected and reported. Recall that $eno = H(t \oplus m)$, $P_{(eno, v_i)} = P_{v_i}(eno)$ and $vno = H(t \oplus m \oplus P_{(eno, v_i)})$. As $H$ is a collision resistant hash function, a valid $vno$ value for the same $<$message, sender, $t>$ triplet is unique. Thus, a vehicle can identify identical messages from the same sender via the $vno$ value. Note that for the \textit{current} time frame, that is for the current value of the broadcast secret $t$, this mechanism can only detect messages replayed to the same vehicles.  

\subsection{Protecting Against Internal, Active Adversary}
The pseudonym verification procedure of V2VAuth uses the hardness of quadratic residuosity and prime factorization problems to verify that $x \in \mathbb{QR}_M$, confirming that the sender's share of the secret polynomial was not tampered. But that procedure does not authenticate the value of variable $y$, that was used to compute the pseudonym. Thus, internal active adversary may successfully use the pseudonym of message $m$ for message $m'$. In order to protect from an internal active adversary, we present a simple modification to V2VAuth scheme. Let the collision resistant bivariate polynomial, $P(x,y)$, exhibit a \textit{special} homomorphism, such that if $x_1 + x_2 \in \mathbb{QR}_M$ and $y \in Z_M$, then $P(x_1, y) + P(x_2, y) = P(x_1 + x_2, y)$. The following text explains the modifications to the scheme.
\begin{itemize}
\item First, update the input set of the polynomial, $P(x,y)$. That is the set of quadratic residues, from which the temporary ids, $tid_{v_i}$, are picked, is updated. It is well known that the size of the set of quadratic residues, $s_q(z) \in \mathbb{F}_q$, such that, $x + y = z$, where $x, y, z \in \mathbb{QR}_{\mathbb{F}_q}$, is at least $\dfrac{q+1}{4}$. Therefore, for each $x \in \mathbb{QR}_{p_i}$, where $p_i$ is a prime factor of $M$, there exist $\dfrac{q-3}{4}$ quadratic residues, $y$, such that $(x+y) \in \mathbb{QR}_{p_i}$. Let the set of such $x$ and $y$ quadratic residues modulo $p_i$ be represented by $\mathcal{QR}_{p_i}$, and let the superset formed by all such sets modulo $p_i$, where $p_i \in \mathbb{P}_{c_i}$, be represented by $\mathcal{QR}_{M_i}$. The sets $\mathcal{QR}_{M_i}$ and $\mathbb{P}_{c_i}$ define the vehicle class $c_i$.
\item On receiving message $m$ with pseudonym $P_{(eno, v_1)}$, vehicle $v_2$ performs the following procedure to verify that $y = H(t \oplus m)$ was used to generate $P_{(eno, v_1)}$.
\begin{enumerate}
\item Evaluate its polynomial share at $y = H(t \oplus m)$ to compute $P_{(eno, v_2)} = P_{v_2}(y)$. 
\item Vehicle $v_2$ computes $P' = P_{(eno, v_2)} + P_{(eno, v_1)}$. We know that $tid_{v_1} + tid_{v_2} \in \mathbb{QR}_M$, hence if $y = H(t \oplus m)$ was used to generate $P_{(eno, v_1)}$, then $P' =P(tid_{v_1} + tid_{v_2}, y)$, which outputs a quadratic residue modulo $M$. Finally, $v_2$ performs the quadratic residuosity tests on $P'$. Note that this procedure eliminates the need to perform the quadratic residuosity tests of the basic V2VAuth scheme. 
\end{enumerate}    
\end{itemize} 

\section{Multicast Scheme}
\label{sec:V2VM}
In this section, we present a multicast variant of \textit{V2VAuth}, dubbed \textit{V2VMulticast}. The multicast scheme employs V2VAuth for the communications but isolates the cluster and protects the intra-cluster communications by using a temporal, random cluster membership token. All existing V2V communications schemes require a cluster head for group/cluster of vehicles, hence creating a single point of failure. V2VMulticast is the first scheme to not require a cluster head. Here we only describe the procedure of computing and updating the cluster membership token. 

\subsection{V2VMulticast}
For each cluster, the scheme generates a random cluster membership token, which is updated periodically and can be computed only by the current cluster members. In order to compute the membership token, a vehicle must be in the vicinity of at least $w-1$ cluster members. To form a cluster, at least $e$ vehicles generate and send \textit{cluster forming message} (cfm) to the AA. Vehicle $v_i$ computes its $cfm$ message as: $cfm_i = P_{v_i}(t)$. The AA uses the received $cfm$ messages to compute the cluster id as:
\begin{enumerate} 
\item Use polynomial $P(x, y)$ to compute each vehicle's temporary id, $tid_{v_i}$, from the respective $cfm_i$ value. 
\item Compute $tid_j = \Sigma_{i=1}^e tid_{v_i}$, and generate the cluster id as: $c_j = H(tid_j \oplus t)$.
\end{enumerate} 
For cluster $c_i$, AA generates a cluster specific polynomial, $P_i(x, y)$, with the $x$ and $y$ degree being $w-1$ and $l-1$, respectively. The cluster specific polynomial is used to compute the cluster membership token, $rand_i$. A maximum of $l$ different $rand_i$ values can be securely computed via the shares of $P_i(x, y)$. Thus, AA controls the number of fresh cluster membership tokens that can be securely computed via $P_i(x, y)$, hence enforcing a lifetime for the cluster. At least $w$ vehicles must combine their shares to compute a fresh $rand_i$ value. Fig. 3 illustrates a high level call flow of the procedure followed to add a new vehicle to the cluster and the process to update the cluster membership token. The following text gives a step-wise description. 
\begin{figure}[t!]
    \centering
    \includegraphics[scale=.70]{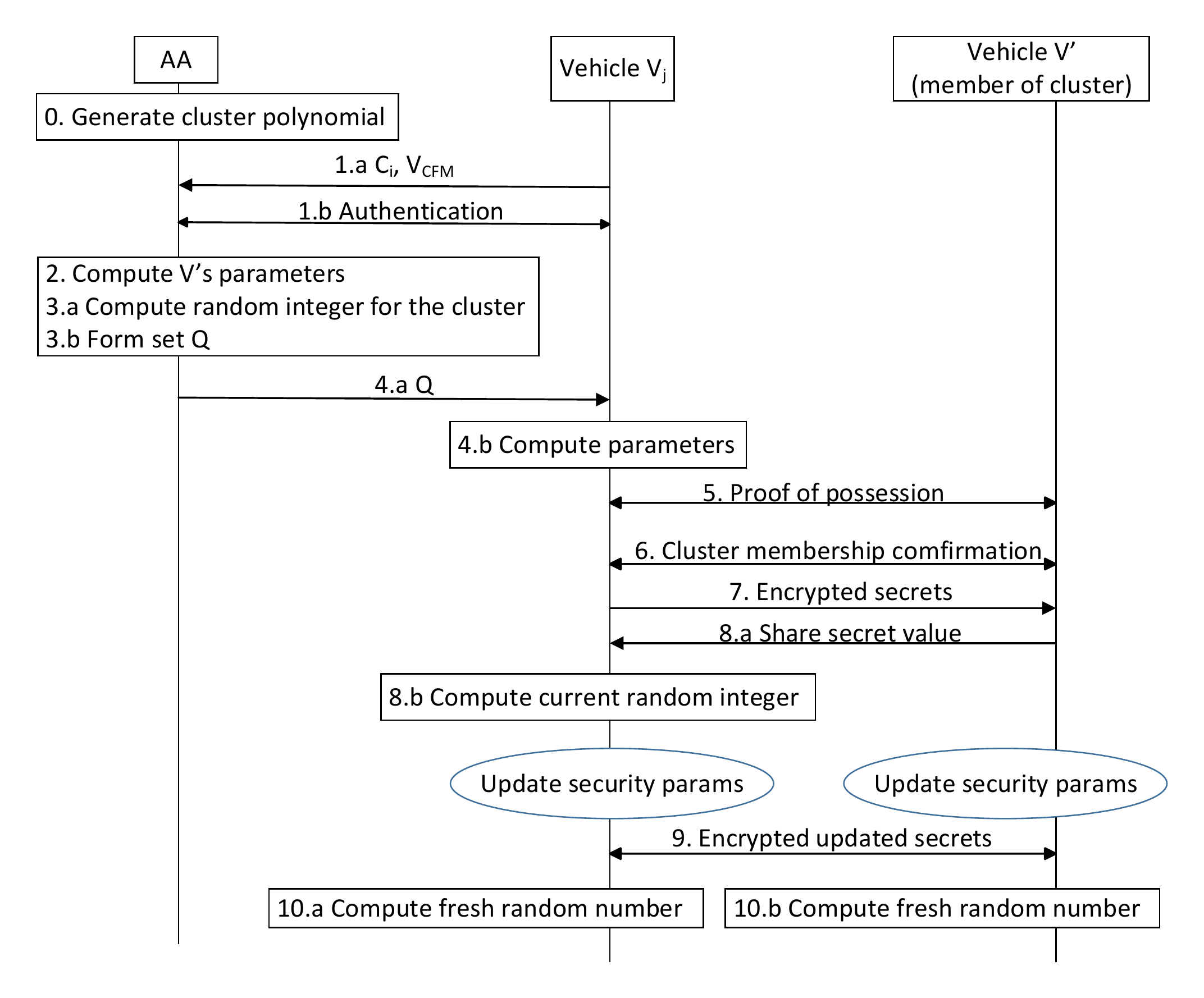}
    \caption{\textit{V2VMulticast} - Adding new member, and generating and updating the cluster membership token}
\end{figure}
\begin{enumerate}
\item Vehicle $v_j$, which is already authenticated for broadcast generates its $cfm_j$ message as: $cfm_j = P_{v_j}(t)$, and sends it to AA along with a request to join cluster $c_i$.
\item AA computes $tid_{v_j}$ from the $cfm_j$ value, and generates the vehicle specific parameters as:
\begin{enumerate}
\item Compute $c_{(i,v_j)} = P_{v_j}(c_i \oplus t)$. 
\item Compute $v_j$'s share of the cluster polynomial as: $P^{i}_{v_j} = P_i(c_{(i,v_j)}, \cdot)$.
\end{enumerate}
\item AA computes $rand_i = P_i(0, c_i \oplus t)$, and constructs set $\mathbb{Q}$ as:
\begin{enumerate}
\item Compute $P^{*}_{(i,v_j)} = c_{(i,v_j)} \oplus P^{i}_{v_j}$.
\item Generate an authenticator, $h_{c_i} = H(rand_i, c_i \oplus t)$, which can be used to confirm the cluster membership of any vehicle. 
\item Compute $xuth_{(i,v_j)}= PRP(rand_i, c_{(i,v_j)})$. Note that only $v_j$ and the AA can compute $c_{(i,v_j)}$. 
\end{enumerate}
\item AA sends $\mathbb{Q} (= \{xuth_{(i,v_j)}, h_{(i,v_j)}, P^{*}_{(i,v_j)}\})$ to $v_j$. 
\item Upon receiving $\mathbb{Q}$, $v_j$ computes the parameters as: 
\begin{enumerate}
\item Compute and save $c_{(i,v_j)} = P_{v_j}(c_i \oplus t)$. 
\item Compute $v_j$'s share of the cluster specific polynomial as: $P^{i}_{v_j} = c_{(i,v_j)} \oplus P^{*}_{(i,v_j)}$
\item Vehicle $v_j$ saves the triplet $(xuth_{(i,v_j)}, h_{c_i}, P^{i}_{v_j})$.
\end{enumerate}
\item Vehicle $v_j$ locates a current member, $v'$, of cluster $c_i$. Similar to Step 7.\ref{Step} of V2VAuth scheme, $v'$ and $v_j$ perform a \textit{Proof of Possession}, confirming that both parties possess the broadcast secrets, $t, \alpha$ and $\gamma$.
\item In order to prove its membership of cluster $c_i$, \textit{v'} sends $h'_{c_i} = H(rand_i, c_i \oplus t)$ to $v_j$.
\item Vehicle $v_j$ compares the received $h'_{c_i}$ value with $h_{c_i}$ value it has saved. If $h_{c_i} = h'_{c_i}$, then $v_j$ sends $xuth_{(i,v_j)}$ to $v'$. 
\item Vehicle $v'$ uses the following procedure to compute the parameters that enable $v_j$ to compute the cluster membership token. 
\begin{enumerate}
\item Decrypt $xuth_{(i,v_j)}$ by using the current $rand_i$ value and generate: $c_{(i,v_j)} = PRP^{-1}(rand_i, xuth_{(i,v_j)})$.
\item Compute $s_{(i,v')} = P^{i}_{v'}(c_i \oplus t)$.
\item Compute $c_{(i,v')} = P_{v'}(c_i \oplus t)$, and $c_{(i,v',v_j)} = c_{(i,v')} \oplus c_{(i,v_j)}$.
\end{enumerate}
\item Vehicle $v'$ sends $\mathbb{E} = \{c_{(i,v',v_j)}, s_{(i,v')}\}$ to $v_j$.
\item On receiving $\mathbb{E}$, $v_j$ computes $rand_i$ as:
\begin{enumerate}
\item $c_{(i,v')} = c_{(i,v',v_j)} \oplus c_{(i,v_j)}$, $s_{(i,v_j)} =P^{i}_{v_j}(c_j \oplus t)$
\item Use $c_{(i,v')}, c_{(i,v_j)}, s_{(i,v')}$ and $s_{(i,v_j)}$ to interpolate the polynomial, $P_i(x,y)$ (employing the bivariate polynomial interpolation from~\cite{Naor[99]}), and compute: $rand_i = P_i(0, c_i \oplus t)$.\vspace{0.5mm}
\end{enumerate}
\textit{Updating cluster specific random integer:} With each update of the value of the broadcast secret $t$, a fresh cluster membership token is generated as:
\item Each cluster member, $v_j$, securely broadcasts its updated $(c_{(i,v_j)}, s_{(i,v_j)})$ value pair. The communication is secured via encryption, using the current $rand_i$ value as the key.
\item Each vehicle that receives at least $w-1$ messages is able to perform the polynomial interpolation and compute the new $rand_i$ value. 
\end{enumerate}

\subsection{Dissolving the Cluster} 
The AA maintains a public record of the cluster ids of all current clusters. The right to dissolve the cluster rests with the \textit{founding} members of the cluster. Recall that the summation, $tid (= \Sigma_{i=1}^e tid_{v_i})$, of the temporary ids of the founding members is used the generate the cluster id. This enables the AA to identify the founding members of the cluster. Every time a founding member of the cluster exits or votes to dissolve the cluster, that event is recorded by the AA. When a threshold number, $e$, of the founding members have either left the cluster or requested to dissolve it, then the AA dissolves the cluster and removes it from the list of current clusters. 

\section{Security and Privacy}
\label{sec:secprf}
In this section, we prove our schemes' security and compare their privacy with the existing schemes. 
\subsection{Broadcast Case (V2VAuth)}
\subsubsection{Message Integrity.}
Each message is accompanied by two values, namely $P_{(eno, v_i)}$ and $vno$. The sender computes the hash of the message, $eno (= H(t \oplus m))$, which is used to compute $vno$ and the pseudonym $P_{(eno, v_i)}$. Thus, if the message gets corrupted then the verification of $vno$ and $P_{(eno, v_i)}$ fails as $H$ is a collision resistant hash function. 
\subsubsection{Source and Pseudonym Authentication.}
The message receiving vehicles first confirm that the sender is an authenticated vehicle by confirming possession of secret parameters, $t$, $\alpha$ and $\gamma$. Recall that each vehicle, $v_i$, is assigned a unique share of the collision resistant polynomial, $P(x,y)$, computed as: $P_{v_i} = P(tid_{v_i}, \cdot)$. As $PRP$ is a secure pseudo random permutation and $P(x,y)$ is collision resistant, $P_{v_i}$ is unique. The \textit{proof of possession} for the broadcast secrets is done via the value of $vno$, which is computed as: $vno = H(t \oplus m \oplus P_{(eno, v_i)})$. So, the computation uses message $m$, the current value of $t$, the collision resistant polynomial $P_{v_i}$ and the secure hash function, $H$. Therefore, the collision resistance of $vno$ values follows trivially.   

The receiver also verifies that the sender did not tamper with its share of the secret polynomial. This is achieved by verifying that the pseudonym provided with the message belongs to $\mathbb{QR}_M$. For each class, $i$ ($1 \leq i \leq u$), except for prime factors $p_i \in \mathbb{P}_{c_i}$, the prime factorization of $M$ is kept secret. AA ensures that values of $cl$ (number of different classes receiving the message) and $b$ ($= |\mathbb{P}_{c_i}|$) are sufficiently large, such that the probability, $2^{-(cl+b)}$, of a fake pseudonym getting through, is negligible. We know that the problem of deciding if an integer is a quadratic residue modulo a RSA modulus $M$ reduces to solving prime factorization of $M$. Hence, as our schemes use a collision resistant bivariate polynomial modulo $M$, forging a fake pseudonym reduces to solving prime factorization of $M$. Finally, the receiver checks whether the sender is a blacklisted party via the output of the identification function, $idnt(P_{(eno,v_i)},tid_{v_j}, eno)$. The $eno$ value is just the hash of the message and the current value of the broadcast secret, $t$. We know that the temporary id, $tid_{v_j}$, is computed using a secure PRP, and therefore leaks no information about the permanent identity of the vehicle. Also, as discussed earlier, $P_{(eno, v_i)}$ does not leak any information about $tid_{v_i}$. Hence, the privacy of the blacklisted parties is preserved.  
\subsubsection{Privacy, Anonymity and Unlinkability.}(i) \textit{Broadcast Scheme:}
The pseudonym of vehicle $v_i$, for message $m$, with the broadcast secret being $t$, is generated as: $P_{(eno, v_i)} = P_{v_i}(H(t \oplus m))$. We know that $P(x,y)$ and $H$ are collision resistant. Hence, for the same message and $t$ value, no two polynomials $P_{v_i}$ and $P_{v_j} (i \neq j)$ have non-negligible probability of a collision. Thus, in order to identify or link different pseudonyms, a polynomial adversary, $\mathcal{A}$, must solve the prime factorization of the RSA modulo, $M$. So, as long as the polynomial, $P(x,y)$, is kept secret, a polynomial adversary has no non-negligible advantage in discovering any relation among different pseudonyms.\\[1.2mm]
(ii) \textit{Multicast Scheme:}
For cluster $c_i$, the cluster membership token, $rand_i$, is computed using a cluster specific polynomial, $P_i(x,y)$. The scheme~\cite{Naor[99]} used to generate the random cluster membership token is a provably secure version of Shamir secret sharing scheme~\cite{Shamir[79]}. If the degree of $y$ in $P_i(x,y)$ is $l-1$, then the scheme allows $l$ secure computations of fresh cluster membership tokens. All intra-cluster communication is secured using the current cluster membership token. Hence, the security and privacy of the multicast scheme follow from the broadcast scheme, V2VAuth.
\subsubsection{Nonrepudiation.}
As argued above, in order to generate a valid, fake pseudonym, a vehicle needs to solve the prime factorization of the RSA modulus $M$. We know that only the AA can identify the sender from the pseudonym and that no party other than the AA can generate quadratic residues modulo $M$. V2VAuth and V2VMulticast allow the receivers to authenticate that $x \in \mathbb{QR}_M$ was used to generate the pseudonym, confirming that the sender did not tamper with its share of the polynomial, $P(x,y)$, and therefore can be identified by the AA. Furthermore, the modified version of V2VAuth allows each vehicle to validate that the correct value of $y$ variable was used to compute the pseudonym. Hence, nonrepudiation is enforced provided that the vehicle's share of the secret polynomial is securely stored.   
\subsubsection{Unforgeability.}
The hash of message $m$, $eno = H(t \oplus m)$, is computed by using a secure hash function. This hash is used to generate the $vno$ value and the pseudonym, $P_{(eno, v_i)}$. Due to $P(x,y)$ and $H$ being collision resistant, if any component of the triplet $<m$, $P_{(eno, v_i)}$, $vno>$ is modified, it gets detected during the authentication process. The only other way to carry out a successful forgery is to mimic the the collision resistant polynomial $P(x,y)$, but as discussed previously, that requires solving the prime factorization problem for the RSA modulus $M$. 
\subsubsection{Proof of Indistinguishability of Pseudonyms.}
We know that $H$ is a secure hash function. Hence, a polynomial adversary has negligible advantage in distinguishing the output of $H$ from a truly random string. The pseudonyms are computed using the collision resistant polynomial $P(x,y)$, evaluated on a secret, pseudorandom input, $tid_{v_i}$. Thus, the advantage of a polynomial adversary in distinguishing any pseudonym from a random string remains negligible.
\subsection{Multicast without Cluster Master (V2VMulticast)}
Security and privacy of broadcast follows from \textit{V2VAuth}. The only new element in the multicast schemes is the cluster membership token. V2VMulticast uses a provably secure variant~\cite{Naor[99]} of Shamir secret sharing scheme to generate the cluster membership token. If $l-1$ is the degree of $y$ in the cluster specific polynomial, $P_i(x,y)$, then upto $l$ membership tokens can be securely computed without leaking any information about the polynomial or the following or previous cluster membership tokens. 
\subsection{Privacy}
All existing V2V communications schemes have a \textit{minimal stable time}, that is the minimum time for which a vehicle must keep the same pseudonym. The minimal stable time ranges from 5 minutes to weeks. Not allowing the vehicles to change their pseudonym for some time weakens the unlinkability of the pseudonyms and exposes the scheme to privacy compromising attacks. In our schemes, vehicles generate a unique pseudonym for each unique message, sender and time frame combination. Hence, our schemes provide true privacy, independent of geographic traffic density and network traffic.  
\begin{table*}[h!]
\begin{centering}
\begin{tabular}{p{3.2cm} p{1.2cm} p{1.2cm} p{1.2cm} p{1.2cm} p{1.2cm} p{1.2cm}  p{0.6cm}}
\hhline{========} 
\textbf{Scheme} & $T_{eval}$ & $T_p$ & $T_{ep}$ & $T_m$ & $T_h$ & $T_{pt}$ & $T_{prp}$\\ 
\hhline{========}
    ECPP~\cite{Lu[08]} & \ding{55} & 3 & \ding{55} & 11 & 1 & \ding{55} & 2 \tabularnewline
    GSB~\cite{Lin[08]} & \ding{55} & 3 & 9 & \ding{55} & 1 & \ding{55} & 1 \tabularnewline
    CPAV~\cite{Vijay[16]} & \ding{55} & 2 & 2 & \ding{55} & 1 & \ding{55} & \ding{55} \tabularnewline
    SCMS~\cite{Whyte[13]} & \ding{55} & \ding{55} & 4 & 6 & 2 & \ding{55} & 4 \tabularnewline
    V2VAuth & 1 & \ding{55} & 1 & \ding{55} & 2 & \ding{55} & 1 \tabularnewline
    V2VMulticast & 3 & \ding{55} & 1 & \ding{55} & 3 & 1 & 2 \tabularnewline
\hhline{========} 
\end{tabular}
\par\end{centering}
\vspace{3mm}\caption{{\small Comparison of the cost of single pseudonym generation and verification $T_{eval}$, $T_p$, $T_{ep}$, $T_m$, $T_h$, $T_{pt}$ and $T_{prp}$ \textit{denote the time complexity of: evaluation of an univariate polynomial of degree $k$ in group $\mathbb{G}$, one pairing operation, exponentiation in $\mathbb{G}$, a single multiplication in $\mathbb{G}$, one hash operation, one polynomial interpolation of an univariate polynomial of degree $k$, and one PRP evaluation, respectively}}}
\end{table*}
\section{Computational Complexity Analysis}
Table 1. summarizes the computational overhead comparison of our schemes with the other pseudonym-based V2V communications schemes. All existing schemes rely on certificate based signature generation and verification for pseudonym generation and authentication, hence mandating the use of costly operations like cryptographic pairings, hefty key management and regular certificate updates. Our schemes utilize more efficient operations such as modular exponentiation and modular multiplication. Compared to the other schemes, our solutions require more hash computations but as hash functions can be evaluated efficiently ($O(1)$, best case), they do not incur much runtime overhead. The SCMS~\cite{Whyte[13]} scheme, which is currently being adopted for the DSRC (Dedicated Short Range Communications) standard, requires four PRP (AES) computations for each butterfly key pair generation. We know that AES-256 is more than 11 times slower than SHA-256~\cite{Crypto[16]}. Hence, SCMS has a significant runtime overhead due to the PRP evaluations.

\section{Future Work}
While designing the schemes, our main emphasis was on achieving anonymity without using certificates and the traditional signature based solutions. Our schemes provide multiple advantages when compared to the existing solutions but assume honesty and privacy guarantee for the \textit{AA}. In the presented schemes, the \textit{AA} possesses the secret polynomial $P(x,y)$, and the key $k$, allowing it to identify messages from the same vehicle. Hence, splitting the \textit{AA} into different parties with unique responsibilities, such that no single party can identify the vehicles, is required to enhance privacy. Our schemes generate a unique pseudonym per new message, which might not be required in some settings. Thus, relaxed versions of the schemes, allowing vehicles to use the same pseudonyms for some fixed amount of time would improve the applicability. Another interesting direction would be to explore information theoretically secure solutions, leading to post-quantum secure V2V communications schemes.   

\section{Conclusion}
\label{sec:cnclsn}
Conflicting security and privacy goals, such as anonymity, non-repudiation and revocation complicate designing efficient V2V communications schemes. Multiple certificate generation and signature verification based schemes have been proposed to achieve security and privacy for V2V communications~\cite{Petit[15]} but these schemes require costly operations such as cryptographic pairings and frequent certificate updates. In this paper, we presented lightweight, non-certificate based V2V communications schemes, which efficiently address the conflicting security and privacy requirements for V2V communications. 

Our schemes address both broadcast and multicast scenarios, and do not require the traditional certificate generation and signature verification. Instead, the presented schemes use provably secure number theoretic results and secret sharing techniques, and except for a central authority, do not make any assumptions about the behavior of third party entities like road side units. The existing pseudonym-based V2V schemes place a limit on the rate at which the pseudonyms may be updated whereas our schemes generate a fresh pseudonym for each unique $<$message, vehicle, time frame$>$ combination. Unlike all other schemes, our multicast scheme does not require a cluster head for a group/cluster of vehicles. Hence, no assumptions about the honesty of the cluster head is required, instead our scheme assumes availability of a central authority for cluster members' registration and provisioning. The time complexity of our schemes is comparable to other pseudonym-based V2V communications schemes. 

\bibliographystyle{ieeetr}

\bibliography{Ref}

\end{document}